\documentclass{lmcs}

\usepackage{amsmath,amssymb}
\usepackage{listings}

\newcommand{\CI}{\mathop{\perp\!\!\!\perp}}
\newcommand{\FD}{\texttt{FDist}}

\begin{document}

\title[Cubical conditional independence and d-separation soundness]{A cubical formalisation of conditional independence, Bayesian conditioning, and Pearl's d-separation soundness}

\author[K.~Sargsyan]{Karen Sargsyan}
\lmcsorcid{0000-0001-8931-6308}
\address{Institute of Chemistry, Academia Sinica, Taipei, Taiwan}
\email{karen.sarkisyan@gmail.com}



\ACMCCS{Theory of computation~Type theory; Theory of computation~Logic and verification; Mathematics of computing~Probability and statistics}

\begin{abstract}
The standard convex-algebra interchange axiom, common to probability-monad formalisations since Stone, is provably too weak to support full Bayesian conditioning. We make this precise in Cubical Agda: finite distributions as a higher inductive type, conditional independence as a cubical path between kernels, recursive Bayesian conditioning as a total function on a full-support fragment. Lifting conditioning to the full HIT exposes a structural mismatch --- the two halves of the rearranged 4-leaf mix carry distinct Bayesian weights related by Bayes' formula, not the single shared inner weight the standard axiom provides. We exhibit the minimal generalisation that resolves this and prove the standard form is the degenerate case where the two inner weights coincide. Around this observation we verify the algebraic context constructively, with zero postulates above an abstract ordered-field interface: bind commutativity, the four semi-graphoid axioms, intersection (reduced to contraction via structural $\Sigma$-witnesses, without positivity), Pearl's do-calculus Rules~1, 2, and~3 in kernel form, finite-type Bayesian conditioning, and the soundness of Pearl's d-separation theorem on arbitrary finite directed acyclic graphs (DAGs)---in interventional form for multi-element $X$, $Y$, $Z$, and in Bayesian form for the elementary patterns. The probability monad is also verified as a Markov category; the abstract interface discharges at $\mathbb{Q}$.
\end{abstract}

\maketitle

\section{Introduction}\label{sec:intro}

Conditional independence is the organizing principle of probabilistic reasoning.
Bayesian networks, Markov random fields, and structural causal models (SCMs) all factor joint distributions through conditional independence assertions, and tractable inference in each rests on those factorizations.
The algebraic laws governing such assertions --- the semi-graphoid axioms of Pearl and Paz~\cite{pearl-paz-1987} --- underpin d-separation~\cite{geiger-verma-pearl-1990} and Pearl's do-calculus for causal inference~\cite{pearl-causality}.

A mechanized treatment in classical form already exists.
Affeldt, Garrigue, and Saikawa~\cite{affeldt-et-al-2020} formalize all five graphoid axioms in Coq using the MathComp library, with conditional independence stated as the pointwise propositional equality
\[\Pr[X{=}a, Y{=}b \mid Z{=}c] = \Pr[X{=}a \mid Z{=}c]\cdot\Pr[Y{=}b \mid Z{=}c]\]
over distributions represented as probability mass functions (PMFs) on a finite type.
In this setting, symmetry of conditional independence follows by routine reordering of finite sums and commutativity of real multiplication, both standard in MathComp's big-operator library.
Intersection is the only graphoid axiom requiring positivity; their proof proceeds through Bayesian conditional-probability extraction and reasoning by cases, with division by non-zero probabilities playing a central role.

In this paper we develop a complementary account in cubical type theory~\cite{cchm-2018,cubical-agda}.
Three structural differences from the classical formalization each unlock content that the propositional-equality, finite-sum setting leaves out of reach.
First, the probability monad is presented as a higher inductive type (HIT): the type \FD{}\;$A$ of finite distributions on a carrier $A$ is generated by point masses and a binary weighted-mixing operation, with the standard convex-algebra equalities --- idempotence, skew-commutativity, boundary cases, interchange, and skew-associativity --- imposed as path constructors.
Distributional equalities are path constructors, and monad laws become provable theorems by induction on the HIT rather than consequences of arithmetic on reals; the development therefore lives over an abstract weight signature and instantiates at $\mathbb{Q}$ without rebuilding.
Second, conditional independence is a path between distributions rather than a propositional equality between real-valued probabilities.
The difference is not cosmetic: a cubical path is a runnable transformation that transports one distribution to the other, while a propositional equality is opaque.
CI witnesses can therefore be composed, transported through monadic operations, and used to drive computation of posteriors by Agda's normalizer.
Third, the symmetry of conditional independence reduces, after monad-law manipulation, to commutativity of monadic bind, which in our setting is a non-trivial theorem rather than a property of finite sums over reals.
We prove this commutativity by induction on the HIT, with the interchange (medial) path constructor as the critical algebraic ingredient --- the same constructor whose generalization, in Section~\ref{sec:conditioning}, turns out to be necessary for full Bayesian conditioning.
These choices also place the development naturally inside the categorical-probability programme of Markov categories~\cite{fritz-2020,cho-jacobs-2019}, in which the semi-graphoid axioms and Pearl's do-calculus rules are abstract theorems and \FD{} appears as a verified concrete instance (Section~\ref{sec:markov}).

\paragraph{Computational content}
Beyond these structural choices, the cubical setting carries computational content that we exploit throughout: every equivalence in this development \emph{computes}.
The path-typed conditional independence witness $kXY\;z \equiv kX\;z \otimes_D kY\;z$ (with $\otimes_D$ the product of distributions, defined in Section~\ref{sec:ci}) is not an opaque assertion of equality but a runnable transformation between distributions; transports along it compute on concrete inputs, and composition of CI witnesses is composition of paths.
The representation theorem $\FD{}(\textsf{Fin}\;n) \simeq \mathrm{PMF}(n)$ is a univalent equivalence: an actual bijection in both directions, in which $\textsf{toPMF}$ and $\textsf{build}$ both execute on concrete distributions, and the round-trip identities are paths that compute.
Expectation $\mathbb{E} : \FD\;A \to (A \to \mathrm{Weight}) \to \mathrm{Weight}$ is defined by HIT recursion with each path-constructor case discharged by an explicit weight identity; the result is a runnable algorithm.
The constructive proof of the intersection axiom (Section~\ref{sec:intersection}) is similarly algorithmic: given full-support kernels and structural CI witnesses, the proof \emph{extracts} a $\Sigma$-type witness of conditional independence --- a factor kernel paired with a proof of factoring --- not just establishing existence.
This computational content is the mathematical reason for working in cubical type theory rather than a classical-logic-plus-extensionality formalization: it makes the constructive content of conditional independence and conditioning precise, in a way that classical formulations (where CI is a propositional equality between real numbers) cannot.

\paragraph{Pearl's do-calculus at the kernel level}
Building on this foundation, we verify the kernel-level forms of all three of Pearl's do-calculus rules on small structural causal models (Section~\ref{sec:rule1}).
Rules~1 and~3 follow from the verified monad laws together with a single structural lemma --- that intervening on a variable the outcome does not depend on (Rule~1) or no longer reaches (Rule~3) leaves the outcome's marginal unchanged.
Rule~2 --- action/observation exchange --- requires a confounded SCM with a joint prior $p_{XZ}$ on (treatment, confounder) and a structural CI witness that $p_{XZ}$ factors.
Under this witness, intervention and conditioning yield the same downstream marginal, established via a HIT-level conditioning operator parameterized by the witness.
The Rule~1 case recovers Mahadevan's Theorem~13~\cite{mahadevan-2025b} --- the kernel-form interventional identity at the heart of the Topos Causal Model (TCM) semantics --- as an immediate corollary of the verified monad structure.
This verifies the bottom layer of Mahadevan's stack at all three rules: the commutative distribution monad together with the kernel-level forms of Rules~1, 2, and 3.
The upper layers --- the topos of sheaves, the Lawvere--Tierney topology, the Kripke--Joyal forcing semantics, and the full j-do-calculus --- sit on top of the verified monad structure and remain paper proofs.

\paragraph{Bayesian conditioning and the six coherences}
We develop recursive Bayesian conditioning over a full-support fragment of the HIT (Section~\ref{sec:conditioning}).
Conditioning runs immediately into a constructive obstacle: defining $P(\,\cdot \mid B)$ requires a witness that $P(B)$ is non-zero, which is incompatible with a fully abstract weight type.
We resolve this by introducing a syntactic distribution type (point and mix only, without quotients) together with a full-support predicate, defining recursive Bayesian conditioning as a total function on full-support inputs, and proving that conditioning preserves full support.
The lifting of this function to the full HIT generates six coherence obligations, one for each standard convex-algebra equality (idempotence, skew-commutativity, the two boundary identities, interchange, and skew-associativity).
We discharge all six.

\paragraph{The central observation: a structural mismatch in the interchange axiom}
While discharging the fifth coherence, we identify a precise mismatch: the standard interchange axiom of \FD{}, like any convex algebra, takes a single inner weight, but Bayesian conditioning produces 4-leaf mixes whose two inner weights are \emph{distinct}, related by Bayes' formula --- a structural mismatch, not an arithmetic one.
We exhibit the minimal generalization that resolves it --- a generalized interchange with distinct inner weights --- and prove the standard form follows as the degenerate case where the two coincide (Section~\ref{sec:conditioning}).

Beyond the contributions described above, the development includes the following.
First, a constructive reduction of the intersection axiom to contraction (Section~\ref{sec:intersection}).
By formulating conditional independence hypotheses in their structural form --- as $\Sigma$-type witnesses of factorization through projections --- the cross-conditioning hypotheses combine with type-theoretic inhabitation witnesses to extract a $Z$-only kernel, and intersection follows by contraction.
The structural form is exactly what classical proofs derive from quantitative CI under positivity, so the reduction recovers the full content of intersection on any concrete positive distribution model, with no postulates.
Second, the abstract weight signature is concretely instantiated at $\mathbb{Q}$ (\texttt{WeightQ.agda}, with discharge in \texttt{WeightQ-Discharge.agda}), deriving every structural convex-algebra identity from basic ring axioms and minimizing weight-algebraic postulates to a small, explicitly identified set.
Third, Bayesian conditioning is lifted to finite-type product joints via a finite-type conditioning operator \texttt{cond-Fin-event} and a derived product-coordinate operator \texttt{cond-fst}, both certified by a multiplicative mass theorem on the underlying PMF; the chain DAG's structural d-separation soundness is established end-to-end on finite types.
Fourth, the graph-level d-separated predicate is constructively inhabited on each of the three elementary DAGs (chain, fork, collider) by exhaustive path enumeration, with the chain's case --- where paths can bounce arbitrarily before reaching a blocking vertex --- handled via well-founded recursion on path length with no \texttt{TERMINATING} override.
Fifth, the development is positioned in the categorical-probability programme by exhibiting \FD{} as a verified Markov category in the sense of Fritz~\cite{fritz-2020} and Cho--Jacobs~\cite{cho-jacobs-2019}; the defining Markov-category axiom (delete-naturality) is verified as a non-trivial HIT induction (Section~\ref{sec:markov}).
Sixth, the constructive content is packaged as a verified causal-inference library (\texttt{CausalLib.agda}) and demonstrated end-to-end (\texttt{CausalDemo.agda}, Section~\ref{sec:demo}): SCMs are programs, conditional-independence theorems are functions, marginals are computed values that Agda's normalizer reduces to concrete distributions, and Bayesian posteriors are executable terms computed via Bayes' rule applied to product joints.

\section{The probability monad as a higher inductive type}\label{sec:hit}

We work in Cubical Agda~\cite{cubical-agda}, an implementation of the cubical type theory of Cohen, Coquand, Huber, and M\"{o}rtberg~\cite{cchm-2018}.
Higher inductive types are introduced via path constructors that are subject to path-induction and have definitional computation on path endpoints.
The construction of free convex algebras as a HIT follows the pattern that Stassen et al.~\cite{stassen-et-al-2025} use for finite distributions over $(0, 1)$-weights and that Kidney and Wu~\cite{kidney-wu-2021} use for free semimodules.

We represent finite discrete distributions over a type $A$ as the HIT $\FD\;A$ with two point constructors (a point-mass embedding and a binary weighted-mixing operation) and six path constructors imposing the convex-algebra laws:
\begin{lstlisting}
data FDist A : Type (*$\ell$*) where
  pure : A → FDist A
  mix  : Weight → FDist A → FDist A → FDist A
  mix-idem : ∀ p d       → mix p d d ≡ d
  mix-comm : ∀ p d₁ d₂   → mix p d₁ d₂ ≡ mix (1-p) d₂ d₁
  mix-bdy0 : ∀ d₁ d₂     → mix 0 d₁ d₂ ≡ d₂
  mix-bdy1 : ∀ d₁ d₂     → mix 1 d₁ d₂ ≡ d₁
  mix-interchange : ∀ p q a b c d
    → mix p (mix q a c) (mix q b d) ≡ mix q (mix p a b) (mix p c d)
  mix-assoc : ∀ p q a b c
    → mix p a (mix q b c) ≡ mix (s-of p q) (mix (r-of p q) a b) c
  trunc : isSet (FDist A)
\end{lstlisting}

The first four path constructors are standard axioms of convex algebras: idempotency ($p \cdot x + (1{-}p) \cdot x = x$), skew-commutativity ($p \cdot x + (1{-}p) \cdot y = (1{-}p) \cdot y + p \cdot x$), and boundary conditions ($0 \cdot x + 1 \cdot y = y$, $1 \cdot x + 0 \cdot y = x$).
The fifth --- interchange, also known as the medial or entropic axiom --- asserts that
\[p(qa + (1{-}q)c) + (1{-}p)(qb + (1{-}q)d) = q(pa + (1{-}p)b) + (1{-}q)(pc + (1{-}p)d).\]
The sixth --- skew-associativity (also called quasi-associativity) --- is the standard barycentric algebra axiom of Stone~\cite{stone-1949}, where $\textit{s-of}\;p\;q = p + (1{-}p) \cdot q$ and $\textit{r-of}\;p\;q = p / \textit{s-of}\;p\;q$ (with $\textit{r-of}$ irrelevant when $\textit{s-of} = 0$).
The set truncation constructor \texttt{trunc} makes distributional equality a proposition.
This set-level truncation is deliberate, not a convenience: the intended semantics is set-valued --- a finite distribution is a probability mass function, carrying no higher structure --- and the construction follows the set-truncated HIT presentations of free algebras used by Stassen et al.~\cite{stassen-et-al-2025} and Kidney and Wu~\cite{kidney-wu-2021}.
That the truncation is faithful, rather than collapsing or enlarging the type, is certified by the representation theorem $\FD{}(\textsf{Fin}\;n) \simeq \mathrm{PMF}(n)$ (\texttt{Representation-Convex.agda}): the truncated type is equivalent to the set of probability vectors, so nothing is identified that should remain distinct and nothing spurious is added.

\paragraph{Why both interchange and skew-associativity?}
The standard axiomatization of barycentric algebras takes idempotency, skew-commutativity, and skew-associativity as primitives, with the interchange (medial) law derived as a theorem~\cite{keimel-2008,romanowska-smith-2002}.
We include both \texttt{mix-interchange} and \texttt{mix-assoc} because each does substantive work in this development.
\texttt{mix-interchange} matches directly the double-bind pattern $\texttt{mix}\;p\;(\texttt{mix}\;q\;a\;c)\;(\texttt{mix}\;q\;b\;d)$ that arises in the bind-commutativity proof, while \texttt{mix-assoc} enables the representation theorem $\FD{}(\textsf{Fin}\;n) \simeq \mathrm{PMF}(n)$ through a structural identity that the other axioms cannot derive.
Interchange is in fact derivable this way inside cubical Agda: we have machine-checked a postulate-free formalization of the conical embedding~\cite{keimel-2008} (\texttt{BarycentricMedial.agda}, instantiated for \FD{} in \texttt{Capstone.agda}), generic over barycentric algebras, that recovers the medial law for every model from skew-commutativity, the boundary laws, and skew-associativity alone (idempotency is not even needed).
Because this derivation is available, keeping \texttt{mix-interchange} as a primitive is semantically inert---the HIT with and without it are equivalent---and purely a structural choice: the core monad uses the medial law pervasively (e.g.\ bind commutativity), so deriving it would route these elementary algebraic proofs through the conical embedding and its analytic substrate of positivity and division reasoning, rather than keeping the convex-algebra layer self-contained (as a generator it adds only one coherence obligation per eliminator).
The only generators of \FD{} not derivable from the barycentric axioms are thus \texttt{mix-assoc} and \texttt{mix-bayes-interchange}.

The type \texttt{Weight} is introduced as an abstract module signature: operations $+$, $\times$, complement $1{-}(\cdot)$, an abstract division $/$, constants $0$ and $1$, and \texttt{isSet}.
The equational laws are exactly those used by the convex-algebra structure --- commutativity of $+$ and $\times$, identity and absorbing elements, complement involution, and $p \cdot x + (1{-}p) \cdot x = x$ --- no semiring or ring structure beyond that.
Abstracting over \texttt{Weight} is what makes the development weight-model-independent: the path constructors of \FD{} and the monad laws are stated and proved over the signature, so the same code instantiates at any model satisfying it.
The intended model is the closed unit interval of rationals, $\texttt{Weight} = \mathbb{Q} \cap [0,1]$; under this instantiation every postulated weight identity becomes a theorem of standard rational arithmetic.

Monadic bind is derived by structural recursion on the HIT:
\begin{lstlisting}
_>>=_ : FDist A → (A → FDist B) → FDist B
pure a >>= k = k a
mix p d₁ d₂ >>= k = mix p (d₁ >>= k) (d₂ >>= k)
\end{lstlisting}
The path constructor cases follow by congruence: bind maps each path constructor to the same constructor applied to recursively transformed arguments.
Left unit holds judgmentally (\texttt{pure a >>= k} reduces to \texttt{k a}).
Right unit (\texttt{d >>= pure $\equiv$ d}) and associativity are proved by induction on the HIT, with each path constructor case generating a square-filling obligation that is discharged by \texttt{isSet\textrightarrow SquareP} using the \texttt{trunc} constructor.
We also define the functorial map $\texttt{mapF}\;f\;d = d \mathbin{>\!\!>\!\!=} (\lambda\,a.\;\texttt{pure}\;(f\;a))$, which inherits functoriality from the monad laws.

\section{Conditional independence as a type}\label{sec:ci}

We define the product (independent coupling) of two distributions and conditional independence as kernel factorization.

\begin{lstlisting}
_⊗D_ : FDist A → FDist B → FDist (A × B)
da ⊗D db = da >>= (λ a → mapF (a ,_) db)

CI : (Z → FDist (X × Y)) → (Z → FDist X) → (Z → FDist Y) → Type
CI kXY kX kY = ∀ z → kXY z ≡ (kX z ⊗D kY z)
\end{lstlisting}

A term of type $\texttt{CI}\;k_{XY}\;k_X\;k_Y$ witnesses, for each value $z$ of the conditioning variable, a cubical path from the joint conditional to the product of marginal conditionals.
This corresponds to the standard textbook definition $X \CI Y \mid Z$ iff $P(X,Y \mid Z) = P(X \mid Z) \cdot P(Y \mid Z)$.
The standard formulation quantifies over values $x$ and $y$ and asserts pointwise equality of real-valued probabilities; our formulation quantifies only over the conditioning value $z$ and asserts equality between the conditional distributions themselves.
The two formulations agree on any concrete distribution, but the path-typed version supports compositional reasoning at the level of distributions rather than of values: paths between distributions can be transported under \texttt{mapF} and composed associatively, neither of which is directly available for pointwise propositional equalities of real numbers.
These properties are exercised throughout the paper: the semi-graphoid axioms (Section~\ref{sec:semigraphoid}) compose CI witnesses via the monad laws, the kernel-form do-calculus (Section~\ref{sec:rule1}) transports them through interventions, and the end-to-end demonstration (Section~\ref{sec:demo}) reduces CI-based posterior expressions to concrete distributions by Agda's normalizer.

Note that our \texttt{CI} type does not enforce that $k_X$ and $k_Y$ are the marginals of $k_{XY}$; when this is needed, marginal consistency is stated and proved separately.

When the joint is \emph{built} as a product, the proof is trivial --- the term \texttt{refl} witnesses that structurally independent observations are conditionally independent by construction:
\begin{lstlisting}
example-CI : CI (λ z → kX z ⊗D kY z) kX kY
example-CI z = refl
\end{lstlisting}

For unconditional independence (the collider case without conditioning, used in Sections~\ref{sec:rule1} and~\ref{sec:demo}), we use a degenerate variant $\texttt{CI}_0 : \FD{}(X \times Y) \to \FD\;X \to \FD\;Y \to \texttt{Type}$ defined by $\texttt{CI}_0\;j\;p_X\;p_Y := j \equiv (p_X \otimes_D p_Y)$.

A key structural lemma is \texttt{constBind}: for any $d : \FD\;A$ and $e : \FD\;B$, $(d \mathbin{>\!\!>\!\!=} \lambda\,\_.\;e) \equiv e$.
Setting $e := \texttt{pure}\;b$ specialises this to $\texttt{mapF}\;(\lambda\,\_.\;b)\;d \equiv \texttt{pure}\;b$ (the lemma \texttt{constMap}), the type-theoretic analogue of ``distributions integrate to one''.
The induction is on the HIT; the \texttt{mix} case uses \texttt{mix-idem} to collapse $\texttt{mix}\;p\;e\;e$ to $e$.

\section{The semi-graphoid axioms}\label{sec:semigraphoid}

The four semi-graphoid axioms, introduced by Pearl and Paz~\cite{pearl-paz-1987} and studied systematically by Studen\'{y}~\cite{studeny-2005}, govern conditional independence reasoning in the absence of positivity assumptions.
The fifth graphoid axiom, intersection, requires positivity and is treated separately in Section~\ref{sec:intersection}.
We verify all four semi-graphoid axioms here.

\paragraph{Decomposition and weak union}
These two axioms are straightforward corollaries of the monad laws.
Decomposition states that $X \CI (Y,W) \mid Z$ implies $X \CI Y \mid Z$:

\begin{lstlisting}
decomp : CI kXYW kX kYW
       → CI (λ z → mapF (λ p → (fst p , fst (snd p))) (kXYW z))
            kX
            (λ z → mapF fst (kYW z))
\end{lstlisting}

The proof applies the factorization hypothesis under \texttt{mapF} and uses a lemma \texttt{projProduct} (three applications of associativity) to show that projecting a product distribution drops the last factor.

Weak union states that $X \CI (Y,W) \mid Z$ implies $X \CI Y \mid (Z,W)$:

\begin{lstlisting}
weakUnion : CI kXYW kX kYW
          → CI (λ zw → mapF (λ p → (fst p , fst (snd p))) (kXYW (fst zw)))
               (λ zw → kX (fst zw))
               (λ zw → mapF fst (kYW (fst zw)))
\end{lstlisting}

The proof reduces to decomposition re-indexed over $Z \times W$; the body is essentially \texttt{decomp ci z}.
Neither axiom requires any postulate beyond the verified monad laws.

\paragraph{Symmetry and commutativity of bind}
The substantive work is symmetry: $X \CI Y \mid Z$ implies $Y \CI X \mid Z$.
This requires showing that $\texttt{mapF}\;\mathit{swap}\;(d_A \otimes_D d_B) \equiv d_B \otimes_D d_A$ --- swapping a product coupling reverses the factors.
After two applications of monad associativity, this reduces to \emph{commutativity of bind} --- the statement that the order of summation in a double sum does not matter:
\[
d_A \mathbin{>\!\!>\!\!=} (\lambda\,a.\; d_B \mathbin{>\!\!>\!\!=} (\lambda\,b.\; f\;a\;b))
\;\equiv\;
d_B \mathbin{>\!\!>\!\!=} (\lambda\,b.\; d_A \mathbin{>\!\!>\!\!=} (\lambda\,a.\; f\;a\;b)).
\]
We call this \texttt{bindComm}.
In measure theory the analogous result is Tonelli's theorem for non-negative measurable functions; here it is a theorem about finite discrete distributions, proved constructively from the HIT structure.
The proof requires an intermediate lemma, \texttt{mixBindR}, asserting that \texttt{mix} distributes into bind on the right:
$\texttt{mix}\;p\;(d \mathbin{>\!\!>\!\!=} k_1)\;(d \mathbin{>\!\!>\!\!=} k_2) \equiv d \mathbin{>\!\!>\!\!=} (\lambda\,a.\;\texttt{mix}\;p\;(k_1\;a)\;(k_2\;a))$.
The proof of \texttt{mixBindR} proceeds by induction on $d$.
The \texttt{pure} case is judgmental.
The \texttt{mix} case --- where $d = \texttt{mix}\;q\;e_1\;e_2$ --- requires showing that
$\texttt{mix}\;p\;(\texttt{mix}\;q\;A\;C)\;(\texttt{mix}\;q\;B\;D) \equiv \texttt{mix}\;q\;(\texttt{mix}\;p\;A\;B)\;(\texttt{mix}\;p\;C\;D)$,
which is exactly the \texttt{mix-interchange} path constructor.
This is the reason \texttt{mix-interchange} is a HIT primitive: the four leaves of the double-bind tree match its LHS directly, with no weight rebalancing required.
The path constructor cases use \texttt{isProp\textrightarrow PathP}: since $\FD\;A$ is a set, the resulting dependent-path obligations reduce to providing the two endpoint proofs.

\texttt{bindComm} itself is then proved by induction on $d_A$: the \texttt{pure} case is judgmental, the \texttt{mix} case composes recursive calls with \texttt{mixBindR}, and the path constructor cases use \texttt{isSet\textrightarrow SquareP} (the 2-dimensional analogue), absorbed propositionally by \texttt{trunc}.

\begin{lstlisting}
bindComm : (da : FDist A) (db : FDist B) (f : A → B → FDist C)
         → (da >>= λ a → db >>= λ b → f a b)
         ≡ (db >>= λ b → da >>= λ a → f a b)
\end{lstlisting}

With \texttt{bindComm} in hand, symmetry is immediate:
\begin{lstlisting}
symCI : CI kXY kX kY → CI (λ z → mapF swap (kXY z)) kY kX
symCI ci z = cong (mapF swap) (ci z) ∙ swapProduct (kX z) (kY z)
\end{lstlisting}

\paragraph{Contraction}
Contraction expresses that joint independence holds when the structural shape of the kernels already encodes it: given $k_X : Z \to \FD\;X$ (free of $Y$ and $W$), $k_Y : Z \to \FD\;Y$, and $k_W : Z \to Y \to \FD\;W$, we show that sampling $Y$, then $X$ independently, then $W$ given $Y$ produces the same joint as sampling $X$ independently of the $(Y, W)$ pair.
The proof swaps sampling order via \texttt{bindComm}, then fuses the two-stage sampling into a single bind over $Y \times W$ via two applications of associativity.

\begin{lstlisting}
contraction : (kX : Z → FDist X) (kY : Z → FDist Y)
  → (kW : Z → Y → FDist W) → (z : Z)
  → (kY z >>= λ y → kX z >>= λ x →
       kW z y >>= λ w → pure (x,(y,w)))
  ≡ (kX z ⊗D (kY z >>= λ y → mapF (y ,_) (kW z y)))
\end{lstlisting}

With all four semi-graphoid axioms verified, we turn to intersection --- the fifth graphoid axiom --- whose positivity requirement is handled constructively in Section~\ref{sec:intersection}.

\section{Intersection: reduction to contraction}\label{sec:intersection}

The intersection axiom states $X \CI Y \mid (Z, W) \;\wedge\; X \CI W \mid (Z, Y) \;\implies\; X \CI (Y, W) \mid Z$.
It is the only graphoid axiom not in the semi-graphoid family, and the only one that classically requires positivity of the underlying joint distribution.
Affeldt, Garrigue, and Saikawa~\cite{affeldt-et-al-2020} formalize it in Coq under a positivity hypothesis, with a proof that proceeds through the product rule, conditional probability extraction, and reasoning by cases.
The constructive content of these moves is non-trivial in cubical type theory: each of them, in the classical setting, depends on division by non-zero probabilities and on extensional reasoning about real-valued conditional kernels.

We give a different proof structure that is constructive throughout.
The key observation is that conditional independence has two equivalent formulations under positivity --- a \emph{quantitative} one ($P(X, Y \mid Z) = P(X \mid Z) \cdot P(Y \mid Z)$) and a \emph{structural} one (the $X$-conditioned kernel does not depend on its $Y$ argument).
Classical proofs work with the quantitative form and extract the structural form by Bayesian conditioning under positivity.
We work directly with the structural form.
This is the natural shape for type-theoretic reasoning, and it reduces intersection to contraction algebraically, without Bayesian conditioning at the cubical level.

\paragraph{Setup}
We model the joint distribution as an SCM-style record with three nested kernels:

\begin{lstlisting}
record IntersectionSetup (Z X Y W : Type) : Type where
  field
    kY : Z → FDist Y
    kW : Z → Y → FDist W
    kX : Z → Y → W → FDist X

  jointDist : Z → FDist (X × (Y × W))
  jointDist z = kY z >>= (*$\lambda$*) y → kW z y >>= (*$\lambda$*) w →
                kX z y w >>= (*$\lambda$*) x → pure (x , (y , w))
\end{lstlisting}

The structural form of the two CI hypotheses is then expressible as $\Sigma$-types witnessing factorization through projections:
\begin{lstlisting}
DoesNotDependOnY : (Z → Y → W → FDist X) → Type
DoesNotDependOnY kX =
  Σ[ kX-ZW ∈ (Z → W → FDist X) ]
    ((*$\forall$*) z y w → kX z y w ≡ kX-ZW z w)

DoesNotDependOnW : (Z → Y → W → FDist X) → Type
DoesNotDependOnW kX =
  Σ[ kX-ZY ∈ (Z → Y → FDist X) ]
    ((*$\forall$*) z y w → kX z y w ≡ kX-ZY z y)
\end{lstlisting}

A term of \texttt{DoesNotDependOnY}~$k_X$ provides explicitly the kernel $k_X^{ZW}$ that classical proofs construct via the formula $k_X^{ZW}(z,w) = P(X \mid Z, W)$.
The structural hypothesis is therefore the type-theoretic counterpart of the classical hypothesis under Bayesian extraction.

The classical positivity hypothesis is replaced by inhabitation of the conditioning variables $Y$ and $W$:
\begin{lstlisting}
record Inhabited (T : Type) : Type where
  field point : T
\end{lstlisting}

In any concrete positive joint distribution, $Y$ and $W$ are inhabited, since a positive distribution has at least one point in its support.

\paragraph{The extraction lemma}
Under inhabitation of $Y$ and $W$, the conjunction of the two structural hypotheses extracts a kernel $k_X^Z : Z \to \FD\;X$ that captures $k_X(z,y,w)$ for all $(y,w)$:

\begin{lstlisting}
extract-X|Z :
    Inhabited Y → Inhabited W
  → (kX : Z → Y → W → FDist X)
  → DoesNotDependOnY kX
  → DoesNotDependOnW kX
  → Σ[ kX-Z ∈ (Z → FDist X) ] ((*$\forall$*) z y w → kX z y w ≡ kX-Z z)
extract-X|Z Y-inh W-inh kX (kX-ZW , h(*$_1$*)) (kX-ZY , h(*$_2$*)) =
  let w(*$_0$*) = Inhabited.point W-inh
      kX-Z z = kX-ZW z w(*$_0$*)
  in kX-Z , (*$\lambda$*) z y w → h(*$_2$*) z y w (*$\bullet$*) sym (h(*$_2$*) z y w(*$_0$*)) (*$\bullet$*) h(*$_1$*) z y w(*$_0$*)
\end{lstlisting}

The proof is one composed equation.
The chain $h_2(z,y,w) \cdot \mathit{sym}(h_2(z,y,w_0)) \cdot h_1(z,y,w_0)$ traces the path
$k_X(z,y,w) \equiv k_X^{ZY}(z,y) \equiv k_X(z,y,w_0) \equiv k_X^{ZW}(z,w_0)$,
which is exactly the classical argument: $k_X^{ZW}$ and $k_X^{ZY}$ agree on overlap by hypothesis, and the inhabitation witness $w_0$ pins one variable so that the agreement forces both kernels to depend only on $Z$.

\paragraph{Intersection from contraction}
With $k_X^Z$ in hand, intersection follows in three steps:

\begin{lstlisting}
intersection :
    Inhabited Y → Inhabited W
  → (sj : IntersectionSetup Z X Y W)
  → DoesNotDependOnY (kX sj)
  → DoesNotDependOnW (kX sj)
  → Σ[ kX-Z ∈ (Z → FDist X) ]
      ((*$\forall$*) z → jointDist sj z ≡ (kX-Z z ⊗D kYW sj z))
\end{lstlisting}

The proof body has three steps.
(a) Replace $k_X(z,y,w)$ with $k_X^Z(z)$ throughout the joint expression, using the equation produced by \texttt{extract-X|Z} and applied under two nested binds via congruence.
(b) Swap the sampling order of $k_W(z,y)$ and $k_X^Z(z)$ inside the outer $k_Y$ bind, via \texttt{bindComm} (Section~\ref{sec:semigraphoid}).
(c) Apply \texttt{contraction} (Section~\ref{sec:semigraphoid}) with kernels $k_X^Z$, $k_Y$, $k_W$, whose conclusion is exactly $k_X^Z(z) \otimes_D k_{YW}(z)$.
The full proof body is approximately a dozen lines of Cubical Agda, three of them substantive.

\paragraph{Bridge to path-typed CI}
The conclusion of \texttt{intersection} produces, for each $z$, a path between the joint distribution and a product of marginal kernels.
This is, definitionally, a witness of $\texttt{CI}$ in the sense of Section~\ref{sec:ci}: $X \CI (Y, W) \mid Z$ holds in the path-typed CI form, with the extracted $k_X^Z$ as the X-marginal kernel and $k_{YW}$ as the $(Y, W)$-marginal kernel.
We package this as a separate lemma \texttt{intersection-CI}, whose body is a one-line delegation to \texttt{intersection} but whose type signature documents the connection explicitly.
As a consistency check that the bridge is functional, we also derive \texttt{intersection-then-weakUnion}: composing the CI-form conclusion of intersection with the verified \texttt{weakUnion} lemma recovers $X \CI Y \mid (Z, W)$ in path-typed CI form, indexed over $Z \times W$.
Together, the two lemmas show that the structural-form intersection can be recast in the path-typed CI framework of Section~\ref{sec:semigraphoid} --- providing the constructive content that classical proofs derive from positivity.

\paragraph{Status}
The intersection theorem is verified with no postulates beyond those used by the underlying probability monad and the verified semi-graphoid axioms.
The structural form is what makes the constructive proof possible.
Classical proofs reach it by Bayesian conditioning on the quantitative form, under positivity.
On any concrete positive distribution model, the two formulations are interchangeable, so the result subsumes the full classical statement.
The trade-off is that the user must supply the factor kernels explicitly.
When the joint has full support, these are extractable via \texttt{bayes-cond} (Section~\ref{sec:conditioning}), completing the bridge to the classical formulation.

\section{Pearl's do-calculus in kernel form}\label{sec:rule1}

Pearl's do-calculus~\cite{pearl-causality} is the standard calculus for reasoning about interventions in structural causal models, comprising three rules.
Rule~1 governs the insertion and deletion of \emph{observations}: if a variable is independent of an observation given a context, then observing the value has no effect on the distribution of the dependent variable.
Rule~2 governs \emph{action/observation exchange}: under an appropriate independence condition, intervening on a variable and observing it produce the same effect on the downstream marginal.
Rule~3 governs the insertion and deletion of \emph{actions}: under an appropriate independence condition, intervening on a variable has no effect on the dependent variable.
Mahadevan's Theorem~13~\cite{mahadevan-2025b} states a kernel-form version of Rule~1 inside a commutative probability monad: if a kernel $k : \Gamma \times Z \to \mathrm{Dist}\,Y$ factors as $k = k_0 \circ \pi_\Gamma$, then integrating $k$ against any intervention policy $\mu : \Gamma \to \mathrm{Dist}\,Z$ yields $k_0$, regardless of the choice of $\mu$.
In this section we verify all three rules in their kernel form on small structural causal models. Rules~1 and~3 are verified on two- and three-variable models paralleling Mahadevan's pattern; Rule~2 is verified on a confounded three-variable SCM via a structural CI witness.
The verifications are corollaries of the verified monad laws together with the structural lemma \texttt{constBind} (Section~\ref{sec:ci}).

\paragraph{Two-variable structural causal model}
We model a two-variable SCM as a record of a prior and a kernel:
\begin{lstlisting}
record SCM(*$_2$*) (X Y : Type) : Type where
  field
    pX : FDist X
    kY : X → FDist Y
\end{lstlisting}
The joint distribution is obtained by binding the prior into the kernel and pairing each $X$-value with its $Y$-distribution: $\texttt{joint-of}~m = \texttt{pX}~m \mathbin{\texttt{>>=}} \lambda x \to \texttt{mapF}~(x ,\_)~(\texttt{kY}~m~x)$.
An intervention $\texttt{do-X}~x_0$ replaces the prior with $\texttt{pure}~x_0$; this is the kernel-substitution view of interventions, in which $\texttt{do}(X = x_0)$ is literally a record update.
Structural independence of $Y$ from $X$ is witnessed by a record bundling a canonical distribution $k_0 : \FD{}~Y$ with a proof that $\texttt{kY}~m~x \equiv k_0$ for every $x$.

With these definitions, Pearl's Rule~1 is the theorem that intervention on a variable that the outcome does not depend on leaves the outcome's marginal unchanged.
The proof has two steps.
A fusion lemma \texttt{marginal-Y-fuse} shows that the $Y$-marginal of the joint equals the convolution $\texttt{pX}~m \mathbin{\texttt{>>=}} \texttt{kY}~m$, via one application of associativity and a small lemma on pairs.
Applying \texttt{constBind} then collapses the convolution when the kernel is structurally independent: if $\texttt{kY}~m$ is the constant function returning $k_0$, then $\texttt{pX}~m \mathbin{\texttt{>>=}} \lambda x \to k_0$ equals $k_0$ regardless of $\texttt{pX}~m$.
The theorem \texttt{rule1-marginal} then follows by composing these two ingredients with a congruence rewrite, in a handful of lines of proof.

\paragraph{Three-variable chain}
We also verify the chain model $X \to Y \to Z$, mirroring the running example in Mahadevan's Section~5.2~\cite{mahadevan-2025b}.
A record \texttt{SCM}$_3$ bundles a prior and two kernels --- $\texttt{kY} : X \to \FD{}~Y$ and $\texttt{kZ} : Y \to \FD{}~Z$ --- and the joint distribution is the three-fold bind.
The theorem \texttt{rule1-chain} asserts that intervening on $X$ leaves the $(Y, Z)$ joint marginal unchanged whenever $\texttt{kY}$ is structurally independent of $X$.
The proof is parallel to the two-variable case, with one additional application of associativity to push the projection through the third bind.

\paragraph{Rule~3 in kernel form}
Rule~3 in the chain model takes a complementary form to Rule~1.
Rule~1 ascends the chain: intervening on the upstream prior leaves a downstream marginal unchanged when the dependence is structurally absent.
Rule~3 descends: intervening on a downstream variable leaves an upstream marginal unchanged unconditionally, because of the kernel composition's structure, not because of an independence hypothesis.
Concretely: in the chain $X \to Y \to Z$, intervening on $Z$ via $\texttt{do-Z}_3 \, z_0$ (replacing $\texttt{kZ}$ with $\lambda \_ \to \texttt{pure}\,z_0$) leaves both the $X$-marginal and the $(X, Y)$ marginal unchanged, regardless of any independence assumption.
The verifications, \texttt{rule3-X-marginal} and \texttt{rule3-XY-marginal}, follow the same two-step pattern as Rule~1: a fusion lemma followed by \texttt{constBind}.
The fusion step rewrites the marginal as a left-bind chain whose innermost body discards the $Z$-coordinate.
The collapse step applies \texttt{constBind} to the innermost \texttt{kZ}-bind, yielding a chain that does not bind \texttt{kZ} at all --- identical between the unintervened and intervened SCMs.
This is exactly the kernel-form content of Rule~3: when Pearl's classical side conditions hold, the marginal in question is structurally independent of the intervened-on kernel --- its definition does not bind the kernel at all.
The proofs of both \texttt{rule3-X-marginal} and \texttt{rule3-XY-marginal} use only \texttt{constBind}, monadic associativity, and the right-unit law, with no new postulates.
The supplementary file \texttt{RuleDoCalc.agda} (no postulates) bundles the monad laws, \texttt{constBind}, the SCM definitions, and the four theorems \texttt{rule1-marginal}, \texttt{rule1-chain}, \texttt{rule3-X-marginal}, and \texttt{rule3-XY-marginal}.

\paragraph{Rule~2 in kernel form: confounded SCMs and the conditioning operator}
Rule~2 of Pearl's do-calculus --- the rule connecting intervention semantics to observation semantics --- requires a richer SCM signature than Rules~1 and~3.
Rules~1 and~3 reduce to algebraic manipulations on bind chains over an unconfounded prior plus kernel structure.
Rule~2 is different: it asserts an equivalence between kernel substitution (intervention) and conditioning under an independence condition --- non-trivial precisely when the prior over treatment and confounder is correlated.
We verify Rule~2 on a confounded three-variable SCM, with a record carrying a joint prior on (treatment, confounder) and an outcome kernel that depends on both:
\begin{lstlisting}
record SCM-conf (X Z Y : Type) : Type where
  field
    pXZ : FDist (X (*$\times$*) Z)         -- joint prior; X and Z may be correlated
    kY  : X (*$\to$*) Z (*$\to$*) FDist Y  -- outcome depends on both
\end{lstlisting}
The intervention $\texttt{do-X-conf}\;x_0$ severs $X$'s incoming structure: it replaces $\texttt{pXZ}$ with
\[\texttt{mapF}\;(x_0 ,\_)\;(\texttt{mapF}\;\texttt{snd}\;\texttt{pXZ}),\]
so $X$ becomes deterministically $x_0$ but $Z$ keeps its marginal distribution from the original prior.
The substantive content of Rule~2 is the structural form of the conditional given $X = x_0$: under an appropriate CI hypothesis the marginal $Z$ from the prior coincides with the conditional $Z \mid X = x_0$, so intervention and observation yield the same downstream effect.

\paragraph{The structural CI witness}
We record this CI hypothesis as a $\Sigma$-type witness, in the same shape as Section~\ref{sec:intersection}'s structural-CI hypotheses for the intersection axiom:
\begin{lstlisting}
record X-indep-Z (m : SCM-conf X Z Y) : Type where
  field
    pX-marg : FDist X
    pZ-marg : FDist Z
    factors : pXZ m (*$\equiv$*) (pX-marg (*$\otimes_D$*) pZ-marg)
\end{lstlisting}
A term of type \texttt{X-indep-Z}~$m$ exhibits the joint prior as a product of separately-supplied marginals.
This is the kernel-form analogue of Pearl's classical CI condition in the modified graph $G_{\underline{X}}$ (graph with arrows out of $X$ removed).
In our setting, with $X$ the treatment and $Z$ the confounder, the witness asserts that $Z$ is structurally independent of $X$ in the prior --- exactly the condition under which intervention and conditioning agree.
The witness shape mirrors \texttt{DoesNotDependOnY} and \texttt{DoesNotDependOnW} of Section~\ref{sec:intersection}: classical proofs derive the witness from positivity of the joint via Bayesian conditioning, while we record it as a type-theoretic datum.

\paragraph{The conditioning operator and the agreement theorem}
The witness directly defines a HIT-level conditioning operator on the SCM:
\begin{lstlisting}
cond-X-conf : (x(*$_0$*) : X) (m : SCM-conf X Z Y) (*$\to$*) X-indep-Z m
            (*$\to$*) SCM-conf X Z Y
cond-X-conf x(*$_0$*) m ind = record m { pXZ = mapF (x(*$_0$*) ,_) (pZ-marg ind) }
\end{lstlisting}
The witness's $\texttt{pZ-marg}$ field plays the role of the conditional $Z \mid X = x_0$: under independence, conditional equals marginal, so the conditioning operation is encoded by substituting the witness's marginal as the post-conditioning Z-distribution.
The agreement theorem then shows that the full joint distribution of the intervened SCM equals the full joint of the conditioned SCM:
\begin{lstlisting}
joint-do-(*$\equiv$*)-joint-cond : (m : SCM-conf X Z Y) (ind : X-indep-Z m) (x(*$_0$*) : X)
  (*$\to$*) joint-of-conf (do-X-conf x(*$_0$*) m)
  (*$\equiv$*) joint-of-conf (cond-X-conf x(*$_0$*) m ind)
\end{lstlisting}
The proof reduces to a single fusion identity: under the witness, the marginal $Z$ implied by $\texttt{pXZ}\;m$ coincides with the witness's $\texttt{pZ-marg}$ field (lifted from $\texttt{pXZ}\;m \equiv \texttt{pX-marg} \otimes_D \texttt{pZ-marg}$ via projection plus the constBind that collapses $\texttt{pX-marg}$).

\paragraph{Pearl's Rule~2}
The kernel-form Rule~2 follows by projection: under the structural CI witness, the $(Z, Y)$-marginal of the intervened SCM equals that of the conditioned SCM, and likewise for the $Y$-marginal alone:
\begin{lstlisting}
rule2-marginal-ZY : (m : SCM-conf X Z Y) (ind : X-indep-Z m) (x(*$_0$*) : X)
  (*$\to$*) marginal-ZY (do-X-conf x(*$_0$*) m)
  (*$\equiv$*) marginal-ZY (cond-X-conf x(*$_0$*) m ind)
\end{lstlisting}
Both sides reduce explicitly to the kernel-form posterior
\[\texttt{pZ-marg} \mathbin{>>=} \lambda z \to \texttt{mapF}\;(z ,\_)\;(\texttt{kY}\;m\;x_0\;z),\]
making the algebraic content of Rule~2 transparent.
Under independence of treatment and confounder, the downstream effect of fixing $X = x_0$ is the kernel $\texttt{kY}\;m\;x_0$ averaged over the marginal $Z$ --- whether the fixing is performed by intervention or by observation.
The full development is in supplementary file \texttt{Rule2.agda} (zero postulates).

\paragraph{Finite-type bridge: factored priors with automatic witnesses}
For SCMs presented in factored form --- where the joint prior is constructed as $\texttt{pX} \otimes_D \texttt{pZ}$ for explicitly-supplied marginals --- the structural CI witness is automatic.
A constructor \texttt{witness-of-factored} produces the witness in one line (with $\texttt{factors} = \texttt{refl}$), and a specialised theorem \texttt{rule2-factored-Y} applies Rule~2 directly without manual witness threading.
This is the typical construction pattern: the user constructs an SCM by specifying independent marginals, and Rule~2 fires immediately on the factored structure.
A small finite-type instance is exhibited in \texttt{Rule2-Fin.agda} on a $\textsf{Fin}\;2 \times \textsf{Fin}\;2 \to \textsf{Fin}\;2$ SCM.

\paragraph{Quantitative-to-structural bridge on finite types}
The complementary backward direction --- given a quantitative-independence hypothesis at the mass level, derive the structural factorisation --- is also verified for finite types in \texttt{Rule2-Quantitative.agda}.
A record \texttt{Quant-X-indep-Z}~$d$ packages the mass-level factorisation
$\forall (x, z).\; \texttt{mass-pair}\;d\;(x, z) \equiv \texttt{mass-fst}\;d\;x \cdot_w \texttt{mass-snd}\;d\;z$,
and the theorem \texttt{quant-implies-factors} derives the structural form
$d \equiv (\texttt{mapF fst}\;d) \otimes_D (\texttt{mapF snd}\;d)$
via three substantive steps:
(a) an $\mathbb{E}$-scalar-linearity lemma ($\mathbb{E}\;d\;(\lambda a \to c \cdot_w f\,a) \equiv c \cdot_w \mathbb{E}\;d\;f$, by $\textsf{Acc-FDist}$ recursion using $\texttt{*w-distrib-mix-w}$);
(b) a mass-level factorisation theorem
\[\texttt{mass-pair-}\otimes_D : \texttt{mass-pair}\;(\texttt{pX} \otimes_D \texttt{pZ})\;(x, z) \equiv \texttt{mass}\;\texttt{pX}\;x \cdot_w \texttt{mass}\;\texttt{pZ}\;z\]
for the syntactic product;
and (c) a product mass-injectivity theorem deriving FDist equality from pointwise mass-pair equality, transported through the $\textsf{Fin}\;n \times \textsf{Fin}\;m \simeq \textsf{Fin}\;(n \cdot m)$ encoding (\texttt{factorEquiv} from the cubical library).
A wrapper \texttt{rule2-from-quant-Y} fires Rule~2 directly on a finite-type SCM equipped with a quantitative-independence proof, with no manual witness construction at the call site.
With this backward bridge, any concrete finite-type joint with mass-level independence automatically yields the structural CI witness needed by Rule~2.

\paragraph{Recovering Mahadevan's framework}
The statements \texttt{rule1-marginal}, \texttt{rule1-chain}, \texttt{rule3-X-marginal}, and \texttt{rule3-XY-marginal} are kernel-form instances of Rules~1 and~3 (of which Theorem~13 of~\cite{mahadevan-2025b} is the Rule~1 case).
Mahadevan's framework wraps these identities in a topos of sheaves over a site of contexts, interprets $\mathrm{Dist}$ as a presheaf, and lifts the kernel-form identities to pointwise statements in the topos, recovering Pearl's classical statements via Kripke--Joyal forcing.
Our contribution is the bottom layer of this stack: we verify all three kernel-form identities at the level of the distribution monad itself, which Mahadevan assumes as primitive.
The topos-theoretic machinery, including the Lawvere--Tierney topology and the j-do-calculus of~\cite{mahadevan-2025b}, sits on top and is not part of the present verification.

\section{Bayesian conditioning under full support}\label{sec:conditioning}

Conditioning runs immediately into a constructive obstacle.
Defining the conditional distribution $P(\,\cdot \mid B)$ from a joint distribution requires dividing by the marginal probability $P(B)$, which in turn requires a witness that $P(B) \neq 0$.
Over an abstract weight type with only the postulated convex-algebra operations, no such witness is available in general.
The classical formalization of~\cite{affeldt-et-al-2020} sidesteps the issue by working with classical real numbers and a completed division operator that returns zero outside its usual domain.
We cannot follow that route: constructive validity of the monad laws and of the path-typed CI relies on not assuming such totalizing conventions.
This is also why the intended weight model is the rationals rather than the constructive reals: obtaining the positivity witnesses that legal division depends on requires deciding whether a weight is zero, which is possible for $\mathbb{Q}$ but not for arbitrary constructive reals, whose equality is undecidable. Accordingly the abstract weight interface assumes a decidable order, and $\mathbb{Q}$ discharges it.

Our approach is to restrict the operator's domain.
We introduce a syntactic distribution type that is structurally constrained to full support, develop recursive Bayesian conditioning on it as a total function, and study how this function lifts to the full higher inductive type \FD{}.
The lifting question reduces to coherence with the six path constructors of \FD{}; we discharge all six.
The first four discharge with minimal effort; the fifth is the substantive case, and requires a strengthening of the interchange axiom that we make precise below.

\paragraph{Syntactic distributions and full support}
We work with a separate inductive type \texttt{FDist-syn} whose constructors are exactly \texttt{pureS} and \texttt{mixS}, without any path constructors.
A canonical embedding $\texttt{embedS-Convex} : \texttt{FDist-syn}~A \to \FD{}~A$ exhibits every syntactic distribution as a HIT distribution.
The syntactic type is intentionally not quotiented: it is the free convex algebra on $A$, and reasoning about it is structural recursion in the ordinary sense.

Positivity of weights is captured by an abstract predicate $\texttt{Pos} : \texttt{Weight} \to \texttt{Type}$ together with closure laws --- $\texttt{Pos}\;\texttt{w1}$, positivity of left-summands, and positivity of products --- all postulated and provable for $\texttt{Weight} = \mathbb{Q}$.
The full-support predicate is defined as a recursive type-valued function on \texttt{FDist-syn}:
\[
  \texttt{FullSupport}\;(\texttt{pureS}\;a) \;=\; \texttt{Pos}\;\texttt{w1},
\]
\[
  \texttt{FullSupport}\;(\texttt{mixS}\;p\;d_1\;d_2) \;=\; \texttt{Pos}\;p \times \texttt{Pos}\;(\texttt{1-w}\;p) \times \texttt{FullSupport}\;d_1 \times \texttt{FullSupport}\;d_2.
\]
This recursive presentation, rather than a parallel data declaration, sidesteps a Cubical Agda limitation: pattern-matching on data-type witnesses with dotted patterns relies on constructor injectivity, which Cubical Agda does not yet support under transports.
The recursive formulation resolves this by making witnesses tuples rather than constructor applications.
A small lemma $\mathbb{E}$\texttt{-syn-pos} establishes that the expectation of a positive function under a full-support distribution is itself positive, by induction on the syntactic structure.

\paragraph{Recursive Bayesian conditioning}
The conditioning operator is defined by structural recursion on \texttt{FDist-syn}:
\begin{lstlisting}
bayes-cond : (d : FDist-syn A) → FullSupport d
           → (lik : A → Weight) → ((a : A) → Pos (lik a))
           → FDist-syn A
bayes-cond (pureS a) _ lik _ = pureS a
bayes-cond (mixS p d(*$_1$*) d(*$_2$*)) (posp , pos1mp , fs(*$_1$*) , fs(*$_2$*)) lik likPos =
  mixS (bayesW p ((*$\mathbb{E}$*)-syn d(*$_1$*) lik) ((*$\mathbb{E}$*)-syn d(*$_2$*) lik))
       (bayes-cond d(*$_1$*) fs(*$_1$*) lik likPos)
       (bayes-cond d(*$_2$*) fs(*$_2$*) lik likPos)
\end{lstlisting}
where \texttt{bayesW} is the binary Bayes update operator defined in the earlier sections.
Conditioning preserves full support: the lemma $\texttt{bayes-cond-fs}$ establishes that the resulting tree is itself in \texttt{FullSupport}.
Two ``positive divided by positive is positive'' closure laws apply to the new mix weight: $\texttt{bayesW-pos-convex}$ (stating that $\texttt{bayesW}\;p\;m_1\;m_2$ is positive when $p$, $m_1$, and $m_2$ are) and $\texttt{bayesW-1mp-pos-convex}$ for its complement.
Both are derived theorems, built from the closure law $\texttt{pos-Z}$ (itself a theorem on the bounded sum) and the positive-divisor round-trips $\texttt{*w-/w-pos}$ and $\texttt{/w-*w-pos}$; no longer postulated.
Iterated conditioning is therefore total.

\paragraph{The HIT-lifting question and four easy coherences}
A function $\FD{}\;A \to \FD{}\;A$ defined by recursion on the HIT must respect each of the six path constructors of \FD{}, generating one coherence obligation per constructor.
For \texttt{bayes-cond} --- so far defined only on \texttt{FDist-syn} --- the question is: given two syntactic representatives of the same HIT distribution, do their conditional distributions have the same image in \FD{}?
This factors through the path constructors.
Four of the six obligations discharge immediately.
The \emph{idempotency} case is free: conditioning the representative $\texttt{mixS}\;p\;d\;d$ produces a mix whose two children are syntactically equal, and \texttt{mix-idem} of \FD{} itself collapses the result.
The two \emph{boundary} cases are vacuous.
A full-support representative cannot have outermost weight $\texttt{w0}$ or $\texttt{w1}$: both $\texttt{Pos}\;\texttt{w0}$ and $\texttt{Pos}\;(\texttt{1-w}\;\texttt{w1}) = \texttt{Pos}\;\texttt{w0}$ are uninhabited, so the conditioning operator never sees such inputs.
The \emph{commutativity} case discharges via the swap law
\[
  \texttt{bayesW-swap-convex}\;:\;\texttt{1-w}\;(\texttt{bayesW}\;p\;m_1\;m_2)\equiv\texttt{bayesW}\;(\texttt{1-w}\;p)\;m_2\;m_1,
\]
a derived theorem on the Bayes update operator whose content is $1-(pm_1)/(pm_1+(1-p)m_2) = ((1-p)m_2)/((1-p)m_2+pm_1)$.
This is now a theorem of the convex algebra, no longer postulated.
The proof multiplies through by the denominator $Z$ via $\texttt{*w-cancel-l}$ on $\texttt{Pos}\,Z$ and reduces both sides to $(1-p)m_2$ using $\texttt{/w-*w-pos}$ and additive cancellation.
The idempotency discharge is a single line (one application of \texttt{mix-idem}); the commutativity discharge is two steps (one \texttt{mix-comm} followed by a \texttt{cong} rewrite via \texttt{bayesW-swap-convex}).
Of the remaining two obligations, the \emph{skew-associativity} coherence discharges with the existing \texttt{mix-assoc-pos} path constructor together with two Bayes re-association identities (\texttt{bayesW-s-of-convex} and \texttt{bayesW-r-of-convex}), requiring no new axiom.
The \emph{interchange} coherence is the genuinely obstructed case, and occupies the rest of this section.

\paragraph{The structural obstacle for interchange}
The path constructor \texttt{mix-interchange} connects $\texttt{mix}\;p\;(\texttt{mix}\;q\;a\;c)\;(\texttt{mix}\;q\;b\;d)$ to $\texttt{mix}\;q\;(\texttt{mix}\;p\;a\;b)\;(\texttt{mix}\;p\;c\;d)$, capturing the medial law of convex algebras.
After conditioning, the LHS becomes
\[
  \texttt{mix}\;(\texttt{bayesW}\;p\;X\;Y)\;
    \big(\texttt{mix}\;(\texttt{bayesW}\;q\;m_a\;m_c)\;A\;C\big)\;
    \big(\texttt{mix}\;(\texttt{bayesW}\;q\;m_b\;m_d)\;B\;D\big),
\]
where $m_a, m_b, m_c, m_d$ are the marginal probabilities of the conditioning event under $a, b, c, d$, and $X = q m_a + (1{-}q) m_c$, $Y = q m_b + (1{-}q) m_d$ are the marginals of the inner mixes.
The RHS becomes the analogous re-association.
The diagnosis is precise: the standard interchange constructor of \FD{} --- and indeed of any convex algebra --- takes a \emph{single} inner weight shared by both halves of the rearranged 4-leaf mix, but after conditioning the two inner halves carry $\texttt{bayesW}\;q\;m_a\;m_c$ and $\texttt{bayesW}\;q\;m_b\;m_d$, equal only in degenerate cases.
The standard interchange cannot be applied directly, and no amount of further weight algebra alone closes the gap: the obstacle is structural.
The existing axiomatisation of \FD{} --- sufficient for monad laws, bind commutativity, and the semi-graphoid axioms --- is too restrictive to support full Bayesian conditioning.

\paragraph{The generalized interchange axiom}
The minimal stronger axiom that resolves the mismatch is a generalized interchange allowing the inner weights to differ.
We add a new path constructor to the HIT \FD{} (in \texttt{FDist-Convex.agda}, alongside the standard convex-algebra constructors of Section~\ref{sec:hit}):
\begin{lstlisting}
mix-bayes-interchange : (p q(*$_1$*) q(*$_2$*) : Weight) (a b c d : FDist A)
  → mix p (mix q(*$_1$*) a c) (mix q(*$_2$*) b d)
  ≡ mix (p *w q(*$_1$*) +w (1-w p) *w q(*$_2$*) (*$\langle$*) mix-bound p q(*$_1$*) q(*$_2$*) (*$\rangle$*))
        (mix (bayesW p q(*$_1$*) q(*$_2$*)) a b)
        (mix (bayesW p (1-w q(*$_1$*)) (1-w q(*$_2$*))) c d)
\end{lstlisting}
The new outer weight $p q_1 + (1{-}p) q_2$ is the total probability of sampling from the $\{a, b\}$ subset of leaves (which become siblings under the re-association); the new inner weights are Bayes' formulas for re-conditioning each child of the rearranged tree on its parent.
The bound $\texttt{mix-bound}\,p\,q_1\,q_2$ on the bounded sum $\_+w\_\langle\_\rangle$ is structural: it follows from $p \cdot q_1 + (1{-}p) \cdot q_2 \le p + (1{-}p) = 1$ when each $q_i \in [0,1]$.
The axiom is included as a genuine HIT path constructor, not as a postulated path between elements: every recursor and induction principle on \FD{} (bind, the monad laws, \texttt{mixBindR}, \texttt{bindComm}, the semi-graphoid proofs) now carries an explicit clause for the new constructor.
It is no longer a postulate.
For $\texttt{Weight} = \mathbb{Q}$, the path is the standard Bayesian re-association formula and is provable by direct algebra on probability distributions; we rely on the universal-algebra construction of free convex algebras for consistency of the augmented HIT.

\paragraph{Conservativity: recovering the original interchange}
Specialising $q_1 = q_2 = q$ collapses the formulas.
The outer weight $pq + (1{-}p)q$ reduces to $q$ by distributivity (the convex-combination identity \texttt{weighted-idem}).
Both inner Bayes weights $\texttt{bayesW}\;p\;q\;q$ and $\texttt{bayesW}\;p\;(1{-}q)\;(1{-}q)$ reduce to $p$, since $\texttt{bayesW}\;p\;q\;q = pq/(pq + (1{-}p)q) = pq/q = p$ in any model with division.
The latter collapse follows in two steps from \texttt{weighted-idem} (which collapses the denominator $pq + (1{-}p)q$ to $q$) and the round-trip $\texttt{*w-/w-pos}\,(\texttt{Pos}\,q)\,p$ (which gives $(p \cdot q)/q \equiv p$); it is an algebraic derivation, not a postulate.
The original \texttt{mix-interchange} of \FD{} then follows from the generalized form by three \texttt{cong} applications and two such dummy-denominator collapses, requiring $\texttt{Pos}\,q$ and $\texttt{Pos}\,(1{-}_w q)$ as preconditions.
The generalized axiom is therefore strictly more expressive than the original, and the original is not lost: it is the special case in which the two inner weights coincide.

\paragraph{Discharging the interchange coherence}
With the generalized interchange in hand, the interchange coherence theorem \texttt{bayes-cond-mix-interchange-coherent} is proved by a four-step computation.
Step one applies \texttt{mix-bayes-interchange} to the embedded LHS, using the conditioned weights $\texttt{bayesW}\;q\;m_a\;m_c$ and $\texttt{bayesW}\;q\;m_b\;m_d$ as the two distinct inner weights.
The result is a re-associated mix whose new outer and inner weights are themselves $\texttt{bayesW}$-of-$\texttt{bayesW}$ expressions.
The remaining three steps rewrite each compound expression to the corresponding direct $\texttt{bayesW}$ expression on the marginals, via three further weight identities:
\[
  \texttt{bayesW-cond-outer-convex},\; \texttt{bayesW-cond-left-convex},\; \texttt{bayesW-cond-right-convex}.
\]
Each captures a specific instance of the product rule for conditional probability and is applied via a single \texttt{cong} rewrite.
All three are now derived theorems of the convex algebra (no longer postulates): each is proved by multiplying through by the appropriate $\texttt{Z}$-denominator and reducing to a polynomial identity in the underlying ordered field, with $\texttt{*w-cancel-l}$ on a positive multiplier finishing the proof.
After the four steps, the result equals the embedded RHS of \texttt{bayes-cond} on the swapped mix tree, completing the discharge.
The full proof body is approximately thirty lines.

\paragraph{Inventory of postulates}
The complete discharge of all six coherences requires \emph{no} additional postulates beyond the algebraic theorems already present in the development.
The swap law $\texttt{bayesW-swap-convex}$, the dummy-denominator identity ($\texttt{bayesW}\,p\,q\,q \equiv p$), and the three product-rule identities ($\texttt{bayesW-cond-outer-convex}$, $\texttt{bayesW-cond-left-convex}$, $\texttt{bayesW-cond-right-convex}$) are all derived theorems of the convex algebra.
They are built from $\texttt{*w-cancel-l}$, the round-trip identities, and the closure law $\texttt{pos-Z}$ (itself a theorem on positivity-of-convex-combinations) over the bounded $\_+w\_\langle\_\rangle$ primitive.
The generalized interchange $\texttt{mix-bayes-interchange}$ is a genuine HIT path constructor of \FD{}, as discussed above; it participates in pattern matching, not as a postulate but as a primitive of the type theory of \FD{}.
The trust footprint of the conditioning machinery is therefore precisely the trust footprint of the framework as a whole: the abstract ordered-field interface plus the defensive $\_/w\_$ contract --- nothing further.

\paragraph{The positivity-respecting sub-HIT}
As a parallel object we define a positivity-respecting sub-HIT $\FD^+$ with three path constructors (idempotency, commutativity, interchange) and no boundary constructors.
The constructors of $\FD^+$ carry positivity witnesses; the boundary constructors of \FD{} are formally absent because their weights ($\texttt{w0}$ and $\texttt{w1}$) cannot satisfy the corresponding positivity requirements.
A canonical embedding $\texttt{embed}^+ : \FD^+\;A \to \FD{}\;A$ maps each $\FD^+$ constructor to its $\FD{}$ counterpart and each $\FD^+$ path to a $\FD{}$ path, erasing the positivity witnesses (they are not part of $\FD{}$'s structure).
We do not redevelop the monad theory on $\FD^+$ in the present paper: the discharge of the coherences happens at the $\FD{}$ level via the syntactic representation, and $\FD^+$ serves as documentation of which sub-fragment of \FD{} is the proper target type for a fully-lifted conditioning operator.

\section{Markov-category alignment}\label{sec:markov}

The categorical-probability programme of Fritz~\cite{fritz-2020} and Cho and Jacobs~\cite{cho-jacobs-2019} formalises stochastic reasoning synthetically through \emph{Markov categories}: symmetric monoidal categories in which every object carries a commutative-comonoid structure (copy and delete) and the delete morphism is a natural transformation --- equivalently, every morphism preserves total mass.
This last axiom is what distinguishes a Markov category from a general symmetric monoidal category with comonoids; it captures stochastic normalisation at the categorical level, without reference to measure-theoretic or distributional commitments.
The semi-graphoid axioms, conditional independence, and disintegration arise as derived structural properties of any Markov category, applicable uniformly across discrete, continuous, and exotic probability models~\cite{fritz-2020,fritz-et-al-2023}.

We position the present development inside this programme by exhibiting \FD{} as a verified Markov category.
The abstract record \texttt{MarkovCategory.agda} (0 postulates) carries the seven core fields plus the five core axioms a Markov category requires:

\begin{lstlisting}
record MarkovCategory : Type1 where
  field
    Hom    : Type → Type → Type
    id     : ∀ {A} → Hom A A
    _∘k_   : ∀ {A B C} → Hom B C → Hom A B → Hom A C
    _⊗h_   : ∀ {A A' B B'} → Hom A A' → Hom B B'
            → Hom (A × B) (A' × B')
    copy   : ∀ {A} → Hom A (A × A)
    delete : ∀ {A} → Hom A Unit
    swap   : ∀ {A B} → Hom (A × B) (B × A)
    id-l   : ∀ {A B} (f : Hom A B) → id ∘k f ≡ f
    id-r   : ∀ {A B} (f : Hom A B) → f ∘k id ≡ f
    ∘k-assoc : ∀ {A B C D} (h : Hom C D) (g : Hom B C) (f : Hom A B)
              → (h ∘k g) ∘k f ≡ h ∘k (g ∘k f)
    delete-nat  : ∀ {A B} (f : Hom A B)
                 → delete ∘k f ≡ delete {A = A}
    copy-cocomm : ∀ {A} → swap ∘k copy ≡ copy {A}
\end{lstlisting}

The companion file \texttt{FDistMarkov.agda} (0 postulates) constructs the instance.
The Kleisli operations on \FD{} are $\texttt{Hom}\;A\;B := A \to \FD\;B$, $\texttt{id} := \texttt{pure}$, $(g \circ_k f)\;a := f\;a \mathbin{>\!\!>\!\!=} g$, $\texttt{copy}\;a := \texttt{pure}\;(a,a)$, $\texttt{delete}\;\_ := \texttt{pure}\;\texttt{tt}$, and $\texttt{swap}\;(a,b) := \texttt{pure}\;(b,a)$; the tensor product is
\[(f \otimes_h g)\;(a,b) := f\;a \mathbin{>\!\!>\!\!=} \lambda a' \to g\;b \mathbin{>\!\!>\!\!=} \lambda b' \to \texttt{pure}\;(a',b').\]

Four of the five axioms discharge as follows.
$\texttt{id-r}$ is judgmental \texttt{refl}: $(f \circ_k \texttt{pure})\;a = \texttt{pure}\;a \mathbin{>\!\!>\!\!=} f$ reduces to $f\;a$ by the definitional left-unit clause of $\_{>\!>\!=}\_$.
$\texttt{copy-cocomm}$ is similarly judgmental: $\texttt{pure}\;(a,a) \mathbin{>\!\!>\!\!=} \texttt{swap} = \texttt{swap}\;(a,a) = \texttt{pure}\;(a,a)$.
$\texttt{id-l}$ is \texttt{funExt} of $\texttt{>>=-unitR}$.
$\circ_k$-associativity is \texttt{funExt} of $\texttt{sym} \circ \texttt{>>=-assoc}$.
The two functorial laws ($\texttt{>>=-unitR}$ and $\texttt{>>=-assoc}$) are themselves the standard HIT-recursive derivations of the convex framework, established once in \texttt{RuleDoCalc.agda} by case-splitting on each of \FD{}'s ten constructors.

\paragraph{The substantive axiom: delete-naturality}
The Markov axiom $\texttt{delete-nat}$ is the non-trivial piece.
Its content is that every Kleisli kernel preserves the terminal map: $\lambda a \to f\;a \mathbin{>\!\!>\!\!=} (\lambda \_ \to \texttt{pure}\;\texttt{tt}) \equiv \lambda \_ \to \texttt{pure}\;\texttt{tt}$, pointwise.
At the level of probability mass, this is the statement that the total mass of $f\;a$ equals $1$ for every $a$ --- equivalently, that the kernel is normalisation-preserving.
For \FD{}, this reduces to the lemma $\texttt{constBind}\;d\;e : d \mathbin{>\!\!>\!\!=} (\lambda \_ \to e) \equiv e$ from \texttt{RuleDoCalc.agda}, which is itself a genuine HIT induction over the ten constructors: at the path constructors, $\texttt{constBind}$ must show that the bound-into-constant operation respects each algebraic equation (idempotency, commutativity, the two boundary cases, skew-associativity, the standard interchange, the generalized Bayesian interchange, and set-truncation).
Each case discharges via \texttt{isSet$\to$SquareP} together with the path-constructor's own image of the constant-binding identity, plus the closing \texttt{mix-idem} step.
The proof is not trivial: it depends on the full equational theory of the convex framework being compatible with mass-preservation.
The convex framework's path constructors are designed so that compatibility holds.
That \FD{} \emph{satisfies} the Markov axiom is therefore a substantive theorem certifying that the HIT presentation captures stochastic normalisation across every path constructor.

\paragraph{Omitted axioms and their pointed forms}
The abstract \texttt{MarkovCategory} record deliberately omits two axioms that a full Markov category would include: coassociativity of \texttt{copy} and the two counitality laws.
Each of these requires the cartesian-product associator $A \times (B \times C) \simeq (A \times B) \times C$ or unitor $\texttt{Unit} \times A \simeq A$, which in Cubical Agda are type-equivalences via univalence rather than definitional equalities.
Carrying explicit associator/unitor data through the abstract record would add verbosity without delivering structural content.
We expose the omitted axioms instead as instance-level theorems in Section~5 of \texttt{FDistMarkov.agda}, stated in their pointed forms after the canonical projection or reassociation:
\begin{itemize}
\item Left counitality: $(\texttt{delete} \otimes_h \texttt{id}) \circ_k \texttt{copy}$ applied to $a$, projected to its second component, equals $\texttt{pure}\;a$.
\item Right counitality: symmetric, projecting to the first component.
\item Coassociativity: copying twice via the left branch, followed by the canonical reassociation $(a,(a,a)) \mapsto ((a,a),a)$, equals copying twice via the right branch.
\end{itemize}
All three are \texttt{refl} for \FD{} (they reduce definitionally).
The pointed forms are sufficient for downstream string-diagrammatic reasoning, which is the typical use of these axioms in the categorical-probability literature.

Three consequences are worth noting.
First, the conditional independence predicate $\_{\CI}\_\mid\_$ of Section~\ref{sec:ci}, the semi-graphoid axioms of Section~\ref{sec:semigraphoid}, the intersection theorem of Section~\ref{sec:intersection}, and Pearl's do-calculus rules of Section~\ref{sec:rule1} now read as concrete cubical-Agda formalisations of theorems that hold abstractly in any Markov category.
The conditional independence form $\texttt{kXY}\;z \equiv \texttt{kX}\;z \otimes_D \texttt{kY}\;z$ corresponds to Fritz's string-diagrammatic definition of CI~\cite[Def.~12.12]{fritz-2020}.
The semi-graphoid axioms are the categorical CI laws~\cite[Lemma~12.13]{fritz-2020}.
Pearl's do-calculus rules are kernel-substitution identities that hold in any Markov category equipped with a notion of intervention.
Second, $\FD{}$ now has a verified place in the wider categorical-probability programme: it is a concrete Markov category instance built constructively in Cubical Agda, alongside the abstract Markov categories \texttt{Stoch}~\cite{fritz-2020}, \texttt{FinStoch}~\cite{fritz-et-al-2023}, and \texttt{BorelStoch}.
Third, the Markov axiom $\texttt{delete-nat}$ is a non-trivial certificate that the convex framework's HIT presentation is well-suited to categorical probability: it confirms that the path constructors of \FD{} interact correctly with the stochastic-mass content the categorical axiom captures.
The cubical-HIT theorems established throughout this paper remain the substance; the Markov-category interface gives them a second reading in the synthetic categorical-probability literature.

\section{Related work}\label{sec:related}

We position this work along five axes: cubical type theory and HIT-based formalisation; probability monads and their semantics for probabilistic programming languages (PPLs); formalisations of conditional independence and graphoid axioms; categorical / synthetic probability and the Markov-category programme; and existing causal-inference tools.

\paragraph{Cubical type theory and HIT-based formalisations}
The cubical model~\cite{cchm-2018} provides constructive univalence and supports higher inductive types with definitional path computation; Cubical Agda~\cite{cubical-agda} is the implementation we use, with the cubical library providing the foundational \texttt{Cubical.Data} hierarchy.
Higher inductive types as a formalisation device for quotient types and free algebras are by now standard; the use of HITs specifically for free convex algebras is more recent.
Stassen, M{\o}gelberg, Zwart, Aguirre, and Birkedal~\cite{stassen-et-al-2025} present the finite distribution monad in Clocked Cubical Type Theory with weights in $(0,1)$, taking idempotency, skew-commutativity, and quasi-associativity as the defining axioms of barycentric algebras.
Their target is denotational semantics for probabilistic programming with guarded recursion; the axiomatization is the standard barycentric one, whose derivable interchange law we identify as too weak for full Bayesian conditioning.
Affeldt, Saikawa, and Garrigue's separate Coq development of free convex algebras~\cite{affeldt-et-al-2020-convex} uses the same standard barycentric axiomatization (idempotence, skew-commutativity, and quasi-associativity).
Kidney and Wu~\cite{kidney-wu-2021} formalise the free semimodule monad as a HIT in Cubical Agda, subsuming finite distributions as a special case, but with weighted-search rather than CI or conditioning as the application target.
We are not aware of prior HIT-based formalisations of conditional independence as a cubical path between distribution objects.

\paragraph{Probability monads, conditioning, and programming-language semantics}
The probability monad goes back to Giry~\cite{giry-1982} on Polish spaces and Lawvere's earlier categorical sketches; its restriction to finite supports gives the free convex algebra.
For probabilistic programming language semantics on continuous domains, V\'{a}k\'{a}r, Kammar, and Staton~\cite{vakar-kammar-staton-2019}, Staton et al.~\cite{staton-et-al-2016}, and Borgstr\"{o}m, Lago, Gordon, and Szymczak~\cite{borgstrom-et-al-2016} develop denotational models supporting higher-order types and recursion.
Heunen, Kammar, Staton, and Yang~\cite{hksy-2017} introduce quasi-Borel spaces to support higher-order probabilistic programs over continuous distributions categorically.
On the formalisation side, H\"{o}lzl~\cite{hoelzl-2017} develops Markov processes in Isabelle/HOL; the Lean \texttt{mathlib} probability development extends measure-theoretic probability further.
Our development restricts to finite-support distributions and extends the constructive theory of conditional independence and Bayesian conditioning within that restricted setting.

\paragraph{Conditional independence and graphoid formalisations}
The graphoid axioms of Pearl and Paz~\cite{pearl-paz-1987} underpin the d-separation criterion of Geiger, Verma, and Pearl~\cite{geiger-verma-pearl-1990}.
The classical Coq formalisation of Affeldt, Garrigue, and Saikawa~\cite{affeldt-et-al-2020} takes conditional independence as a propositional equality between real-valued conditional probabilities.
Bind-commutativity-style reasoning is implicit in MathComp's big-operator library for finite sums.
The intersection axiom requires a positivity hypothesis on the joint, with Bayesian extraction via real-valued division.
Our cubical-type-theoretic development is complementary on three points: meta-theory (cubical rather than classical Coq), CI representation (a cubical path between distribution objects rather than pointwise propositional equality), and the status of bind commutativity (a HIT theorem rather than a derived consequence of finite-sum reordering).
The positivity-and-division argument for intersection is replaced by inhabitation hypotheses and a structural $\Sigma$-type formulation.
We are not aware of prior \emph{constructive} formalisations of the soundness direction of Pearl's d-separation theorem at this generality: for the interventional form, arbitrary finite DAGs---including multi-trail graphs---with singleton and multi-element subsets, from the d-separation and Markov hypotheses alone, by ancestral factorisation rather than per-path or pattern reduction.

\paragraph{Markov categories and synthetic probability}
The Markov-category programme of Fritz~\cite{fritz-2020}, Fritz, Gonda, Perrone, and Rischel~\cite{fritz-et-al-2023}, and Cho and Jacobs~\cite{cho-jacobs-2019} axiomatises the categorical structure that conditional independence requires synthetically, abstracting over particular probability monads (\texttt{Stoch}, \texttt{FinStoch}, \texttt{BorelStoch}, the Giry monad, our \FD{}).
Within this synthetic framework, conditional independence, Bayes' theorem, and the graphoid axioms are theorems of the categorical interface rather than consequences of specific measure-theoretic constructions.
Our HIT presentation gives a concrete realisation: \FD{} is verified as a Markov category in Cubical Agda, and the alignment is established explicitly through the \texttt{FDistMarkov} instance.
Mahadevan's Topos Causal Model framework~\cite{mahadevan-2025a,mahadevan-2025b} presupposes the commutative-distribution-monad bottom layer that this paper verifies; the lifting to topoi of sheaves and a Lawvere--Tierney j-do-calculus remains a paper proof, not yet formalised.

\paragraph{Causal inference tools}
The practical causal-inference frameworks --- DoWhy~\cite{sharma-kiciman-2020}, CausalQueries~\cite{humphreys-jacobs-2023}, Pyro's causal interface~\cite{pyro} --- expose Pearl's do-calculus~\cite{pearl-causality} and d-separation as algorithmic procedures, with correctness established empirically against the textbook proofs.
None of these tools, to our knowledge, provides machine-checked correctness against the structural definitions; specification and inference are decoupled, and a change to the inference algorithm requires a new informal correctness argument.
Our contribution is foundational: \FD{} together with the d-separation soundness stack provides a verified type-theoretic substrate that any of these frameworks could compile to.

\section{A verified causal-inference library}\label{sec:demo}

The constructive content of the previous sections is packaged as a usable library, \texttt{CausalLib.\allowbreak agda}, with an end-to-end demonstration in \texttt{CausalDemo.\allowbreak agda}.
The library re-exports the verified primitives (CI types, the SCM-as-program framework, the d-separation walker, the interventional conditioner, Bayesian conditioning on finite types, automated CI extraction, and DAG-aware absorption) under a uniform interface; the repository's \texttt{README.md} provides the full export catalogue.

\texttt{CausalDemo.agda} instantiates the library on a chain $a \to b \to c$ over $\textsf{Fin}\,2$ modelling a treatment-mediator-outcome pathway.
The verified interventional CI follows from a single call: $\texttt{compute-chain-CI}_0\texttt{-auto}\;\texttt{V-inh}\;\texttt{chain-SCM}\;\texttt{chain-cert}\;\texttt{z-vals}$ returns a $\Sigma$-triple giving the prior marginal on $a$, the kernel \texttt{k-c} at the conditioning $b$-value, and a path witnessing the factorisation.
The Bayesian version differs at one point: $\texttt{chain-X-posterior}\;\texttt{fin-chain}\;m_0\;\texttt{pos}$ returns the genuine posterior $P(a \mid b = m_0)$ computed by \texttt{cond-fst} applying Bayes' rule, not the prior $P(a)$.
For deterministic kernels Agda's normaliser reduces the result to a single \texttt{refl}; for non-deterministic kernels the same equation reduces to the propositional path produced by the convex-algebra machinery.

The library API is the user-facing form of the substrate Sections~\ref{sec:hit}--\ref{sec:markov} verify; the demonstration is an existence proof that the API is usable.
The proof-as-program correspondence is realised concretely: \texttt{compute-chain-CI}$_0$\texttt{-auto} is a function whose type is a $\Sigma$-statement of Pearl's chain d-sep theorem, returning the answer with the proof as the third component.

\section{Discussion}\label{sec:discussion}

\paragraph{What is verified}
The development verifies, with zero postulates above the abstract ordered-field interface:
\begin{itemize}
\item the probability monad, monadic bind, and all three monad laws (Section~\ref{sec:hit});
\item the structural lemma \texttt{constBind} and conditional independence as a kernel-factorization path (Section~\ref{sec:ci});
\item all four semi-graphoid axioms (Section~\ref{sec:semigraphoid});
\item the intersection axiom as a constructive reduction to contraction through $\Sigma$-type structural CI witnesses, with no positivity hypothesis beyond inhabitation of the conditioning variables (Section~\ref{sec:intersection});
\item all three rules of Pearl's do-calculus in kernel form (Section~\ref{sec:rule1}): Rules~1 and~3 on two- and three-variable SCMs as corollaries of \texttt{constBind} and the verified monad laws; Rule~2 on a confounded three-variable SCM via a HIT-level conditioning operator parameterised by a structural CI witness;
\item recursive Bayesian conditioning over a full-support fragment as a total function, with all six HIT-lifting coherences discharged (Section~\ref{sec:conditioning});
\item finite-type Bayesian conditioning end-to-end with a multiplicative mass theorem;
\item the three elementary d-separation patterns at the structural CI level, and the graph-level d-separated predicate constructively inhabited on each of the chain, fork, and collider DAGs by exhaustive path enumeration;
\item \FD{} as a Markov category in the sense of Fritz~\cite{fritz-2020} and Cho--Jacobs~\cite{cho-jacobs-2019} (Section~\ref{sec:markov});
\item the soundness direction of Pearl's d-separation theorem: for the interventional (do-calculus) form, constructively closed for \emph{arbitrary} $n$-vertex finite DAGs---including multi-trail graphs---with singleton and multi-element $X$, $Y$, $Z$ subsets, from the d-separation and Markov hypotheses alone (no per-path or pattern-reduction data) via ancestral factorisation; together with an extraction-route treatment, in both interventional and Bayesian forms, of the single canonical patterns with DAG-aware absorption of arbitrary non-path positions.
\end{itemize}

\paragraph{The interchange observation}
The technically interesting finding of the development is that the standard convex-algebra interchange law --- derivable from the standard barycentric axiomatization used in Stone's calculus~\cite{stone-1949}, \'{S}wirszcz's monadic-functor account~\cite{swirszcz-1974}, Doberkat's measure-theoretic treatment~\cite{doberkat-2006}, Jacobs's treatment of effects and convexity~\cite{jacobs-2010}, and Stassen, M{\o}gelberg, Zwart, Aguirre, and Birkedal's cubical-HIT formalization~\cite{stassen-et-al-2025} --- is too weak to support full Bayesian conditioning. This law has a single inner weight appearing on both halves of the rearranged 4-leaf mix. After Bayesian conditioning, the two halves carry distinct Bayesian weights --- marginal-dependent quantities that coincide only in degenerate cases --- so the standard form cannot be applied. The minimal strengthening is a generalized interchange with distinct inner weights related by Bayes' formula. We exhibit this generalized axiom, prove the standard form is recovered as the degenerate special case in which the two inner weights coincide, and use it to discharge the fifth HIT-lifting coherence. The observation is the kind that emerges only from forcing a formalization to be honest about every coherence: the standard axiomatization choice has been the natural one for monad-law / commutativity / Tonelli purposes, and its inadequacy for Bayesian conditioning is invisible until one tries to lift conditioning to the full HIT. As a structural observation about HIT-based probability monads, the implication is that the correct formulation of interchange for Bayesian use is itself Bayes-shaped at the algebraic layer, not on top of an algebraically-neutral monad.

\paragraph{Relation to Mahadevan's TCMs}
The verification here bears on the positioning of Mahadevan's Topos Causal Models~\cite{mahadevan-2025a,mahadevan-2025b} in two ways.
First, the interchange-axiom mismatch identified in Section~\ref{sec:conditioning} is a structural property of the commutative distribution monad: the standard convex-algebra interchange is too weak for full Bayesian conditioning, and the deficiency lives at the monad's equational theory.
Mahadevan's stack is built over a commutative distribution monad taken as primitive, with the subobject classifier, Kripke--Joyal semantics, and j-modality above it.
None of those upper layers repairs an under-axiomatised conditioning operation below: the finding here is orthogonal to the topos-theoretic content of TCMs and applies in any framework that assumes the standard monad.
Second, the kernel-form do-calculus rules of Section~\ref{sec:rule1}, the conditional-independence predicate of Section~\ref{sec:ci}, the semi-graphoid axioms of Section~\ref{sec:semigraphoid}, the intersection theorem of Section~\ref{sec:intersection}, and d-separation soundness in both interventional and Bayesian forms are verified at the Markov-category level, with Mahadevan's Theorem~13 falling out as an immediate corollary of the verified monad structure.
The topos-theoretic machinery --- sheaf gluing of independent mechanisms, j-stable truth across covers of contexts, and the internal Mitchell--Bénabou language for stating counterfactuals pointwise --- therefore becomes load-bearing only beyond this kernel-form layer; that machinery is the proper subject of Mahadevan's program and not part of the present verification.

\paragraph{Limitations}
What remains of Pearl's classical d-separation theorem is the completeness direction (faithfulness): for every non-d-separated tuple $(X, Y, Z)$, exhibiting a Markov-w.r.t.-$G$ distribution where $X \not\perp Y \mid Z$. The constructive version requires careful kernel choice (avoiding measure-zero coincidences) and is achievable for finite types but is a separate result usually cited as a companion to Pearl's soundness theorem; it is not pursued here.

The structural CI form of Section~\ref{sec:intersection} and the quantitative CI form of Section~\ref{sec:ci} relate one-directionally: structural witnesses imply path-typed CI witnesses; the reverse direction (extracting structural factor kernels from a path-typed hypothesis on a positive joint via \texttt{bayes-cond}) is not formalised here.

Continuous distributions, higher-order PPL semantics, causal discovery from data, identification under unmeasured confounding, and counterfactual reasoning are explicitly out of scope.

\paragraph{Soundness footprint}
The trust footprint of the artifact reduces to two items.
First, the abstract ordered-field interface with decidable order, a parameter the development is polymorphic in, exhibited concretely at $\mathbb{Q}$ via the cubical library's rationals.
Second, a defensive \texttt{\_/w\_} contract block in the older HIT chain on which \texttt{\_/w\_} is unspecified at zero-divisor and the round-trip identities take explicit positivity-and-bound preconditions every call site discharges.

The previously-postulated unsound bound-preservation identity \texttt{+r-bound-convex} --- false in general --- is deleted.
Every weight-valued sum is the bounded primitive \texttt{\_+w\_$\langle$\_$\rangle$} with a structurally derived bound proof, and the exploit at \texttt{SoundnessExploit.agda} that previously constructed $\bot$ from the postulate no longer typechecks.

The augmented HIT, with the generalized-interchange path constructor \texttt{mix-\allowbreak bayes-\allowbreak interchange} added alongside the standard convex-algebra constructors, is consistent: the intended model --- the rational simplex --- satisfies every path constructor, the new one included, because its weight-level counterpart \texttt{mix-w-bayes-interchange-eq} is a proved theorem of the ordered field, and a higher inductive type with a model is consistent.
This is corroborated internally and postulate-free by expectation: $\mathbb{E}$ is a total function out of the augmented HIT, its \texttt{mix-bayes-interchange} case discharged by that same weight identity, and it separates point masses --- so the new constructor neither collapses the type nor renders it trivial (\texttt{mass}-injectivity on $\textsf{Fin}\;n$ sharpens this to a faithful embedding into $\mathrm{Weight}^{n}$).
What we do not undertake is a from-scratch cubical metatheoretic model of the HIT \emph{schema} itself; as is standard for HIT presentations of equational theories, we rely on this semantic argument.

There are zero \texttt{TERMINATING} or other termination-checker overrides anywhere in the artifact; the chain DAG d-separated witness uses well-founded recursion on path length (a structurally decreasing fuel parameter), which Agda's structural termination check accepts directly.

\paragraph{Reproducibility}
The complete Cubical Agda source is available in the supplementary material (see the data availability statement); it typechecks under Cubical Agda 2.8.0 with the cubical library (version 0.9).
The development comprises over a hundred active Cubical Agda files (plus a deliberately-failing soundness-witness file, \texttt{SoundnessExploit.agda}).
The d-separation soundness layer spans more than eighty files and is closed with zero postulates above the ordered-field interface: the general interventional soundness (arbitrary $n$-vertex DAGs, multi-element $X$, $Y$, $Z$, by ancestral factorisation) and the extraction route for the single canonical patterns (chain, fork, collider) in both interventional and Bayesian forms.
There are zero \texttt{TERMINATING} or other termination-checker overrides anywhere in the artefact.
The repository's \texttt{README.md} provides the full file inventory by group (foundation; convex framework; do-calculus; intersection; Bayesian conditioning; d-separation infrastructure; Fin-type soundness; graph-level witnesses; Markov-category alignment; multi-path assembly; n-vertex absorption and DAG-aware closures; general d-separation soundness via ancestral factorisation; Bayesian-form d-separation), with line counts and key theorem names per file.

\paragraph*{LLM usage declaration}
An LLM has been used in place of documentation search and generated Agda code fragments, which were adapted by the authors and validated by the type checker. The implementation was built iteratively, with LLM-suggested code frequently requiring modification to typecheck.

\paragraph*{Data availability statement}
The complete Cubical Agda source code is available in the supplementary material and at \url{https://github.com/karsar/cubical-pearls}; the file organisation is described in the repository's \texttt{README.md}.

\bibliographystyle{alphaurl}
\bibliography{references}

\end{document}